\documentclass[journal]{IEEEtran}
\usepackage{textcomp}
\usepackage{url}
\usepackage{verbatim}
\usepackage{etoolbox}
\usepackage{amsmath,amssymb}
\usepackage{graphicx,graphics,color,psfrag}
\usepackage{balance}

\allowdisplaybreaks[4]
\usepackage{accents}
\usepackage{amsthm}
\usepackage{bm}
\usepackage[english]{babel}
\usepackage{multirow}
\usepackage{enumerate}
\usepackage{cases}
\usepackage{stfloats}
\usepackage{dsfont}
\usepackage{color,soul}
\usepackage{amsfonts}
\usepackage{graphicx,amsmath,amssymb}
\usepackage{fancyhdr}
\usepackage{hhline}
\usepackage{graphicx,graphics}
\usepackage{array,color}
\usepackage{amsmath}
\usepackage{float}
\usepackage{amssymb}
\usepackage{amsmath}
\usepackage{bbm}
\usepackage{amsthm}
\usepackage{amsfonts}
\usepackage{graphicx}
\usepackage{epstopdf}
\usepackage{cite}
\usepackage{amsmath,bm}

\usepackage{graphicx}
\usepackage{color}
\usepackage{calc}
\usepackage{diagbox}
\usepackage{subcaption}
\usepackage{booktabs}
\usepackage{multirow}
\usepackage[ruled,vlined]{algorithm2e}
\usepackage{algpseudocode}
\usepackage[normalem]{ulem}

\begin{document}
\bibliographystyle{IEEEtran}

\title{Mixture of Semantics Transmission for Generative AI-Enabled Semantic Communication Systems \thanks{All authors are with Cooperative Medianet Innovation Center and Shanghai Key Laboratory of Digital Media Processing and Transmission, Shanghai Jiao Tong University, Shanghai, 200240, China (e-mail: \{nijunjie2021, wu\_tong, zhiyongchen, xuyin, mxtao, zhangwenjun\}@sjtu.edu.cn). M. Tao and W. Zhang are also with Department of Electronic Engineering, Shanghai Jiao Tong University, Shanghai, 200240, China. }}

\author{Junjie Ni, Tong Wu, Zhiyong Chen, Yin Xu, Meixia Tao, \emph{Fellow, IEEE}, and Wenjun Zhang, \emph{Fellow, IEEE}}

\maketitle

\begin{abstract}
	In this paper, we propose a mixture of semantics (MoS) transmission strategy for wireless semantic communication systems based on generative artificial intelligence (AI). 
	At the transmitter, we divide an image into regions of interest (ROI) and reigons of non-interest (RONI) to extract their semantic information respectively. Semantic information of ROI can be allocated more bandwidth, while RONI can be represented in a compact form for transmission. At the receiver, a diffusion model reconstructs the full image using the received semantic information of ROI and RONI. 
	Compared to existing generative AI-based methods, MoS enables more efficient use of channel resources by balancing visual fidelity and semantic relevance. Experimental results demonstrate that appropriate ROI-RONI allocation is critical. The MoS achieves notable performance gains in peak signal-to-noise ratio (PSNR) of ROI and CLIP score of RONI.
\end{abstract}
\begin{IEEEkeywords}
	Semantic Communications, Generative AI, Mixture of Semantics.
\end{IEEEkeywords}

\section{Introduction}
\label{intro}

Semantic communications has recently emerged as a promising research direction for sixth-generation (6G) wireless communications systems. This paradigm shifts the emphasis from mere symbol-level accuracy to the conveyance of meaning, investigating how effectively the intended information of the transmitter is recovered at the receiver. By leveraging artificial intelligence (AI), particularly generative AI, semantic communications enable the receiver not only to reconstruct but also to generate the desired information.\cite{wu_cddm_2023,lipaper, wu2025icdminterferencecancellationdiffusion,zhang2025semanticsguideddiffusiondeepjoint,highperceptual}


Semantic communications for image transmission predominatly adopts the framework of joint source-channel coding (JSCC). Notable approaches include the deep learning-based end-to-end method DeepJSCC \cite{bourtsoulatze_deep_2019,wu2023vision}, the Swin Transformer based model SwinJSCC \cite{yang_swinjscc_2024}, the state space model-based method MambaJSCC \cite{wu_mambajscc_2024}, and the nonlinear transform-driven model (NTSCC)\cite{ntscc}. These methods utilize the end-to-end neural network architectures to encode image into low-dimensional semantic feature spaces for efficient transmission. Moreover, several studies focus on exploiting the intrinsic properties of the semantic feature space. For example, \cite{adaptive_rate_control} proposes a strategy for dynamically selecting a subset of semantic features for transmission. \cite{choi2025featureimportanceawaredeepjoint} introduces a feature-importance-aware block that partitions and transmits semantic features based on their significance. However, these approaches remain confined to the image semantic space and therefore yield only marginal performnace gains.

In recent years, generative AI techniques for image generation, e.g., diffusion model (DM) \cite{ho2020denoising} and GPT-4o, have driven breakthrough advances. These models can not only accurately parse cross-modal semantics (e.g., jointly understanding text and vision), but also leverage self-attention to capture long-range semantic dependencies across complex inputs. As a result, they achieve orders-of-magnitude improvements in image detail richness, multi-object spatial consistency, and physical plausibility. Thus, this has ushered in an entirely new paradigm for wireless image semantic communications.

Motivated by this, we propose a mixture of semantics (MoS) transmission strategy for wireless semantic communication systems based on generative AI in this paper. Specifically, we leverage the Segment Anything Model (SAM) \cite{kirillov_segment_2023} to partition the transmitted image into a Region of Interest (ROI) and a Region of Non-Interest (RONI), using pixel-level representations for the ROI and lightweight, higher-level semantic descriptions for the RONI. This enables the allocation of more communication bandwidth to accurately transmit the ROI, while the RONI is transmitted in a lightweight textual form. Therefore, we term this joint transmission of heterogeneous semantic modalities as MoS. At the receiver, the complete image is reconstructed using a DM based on the received ROI data and the semantic information of the RONI. Compared to existing image transmission methods based on generative AI, the proposed MoS offers greater flexibility for users to specify, at a pixel level, which areas require high-fidelity transmission and which can be reconstructed based on semantic similarity, thereby enabling more efficient utilization of channel resources. Extensive experimental results demonstrate that appropriately balancing the transmission ratio between ROI and RONI is crucial. The proposed MoS achieves substantial improvements in key evaluation metrics such as peak signal-to-noise ratio (PSNR) of ROI and CLIP score \cite{hessel2022clipscorereferencefreeevaluationmetric} of RONI.

\section{System Model}
\label{sys_model}

\begin{figure*}[htbp]
	\centering
	\includegraphics[width=0.90\textwidth]{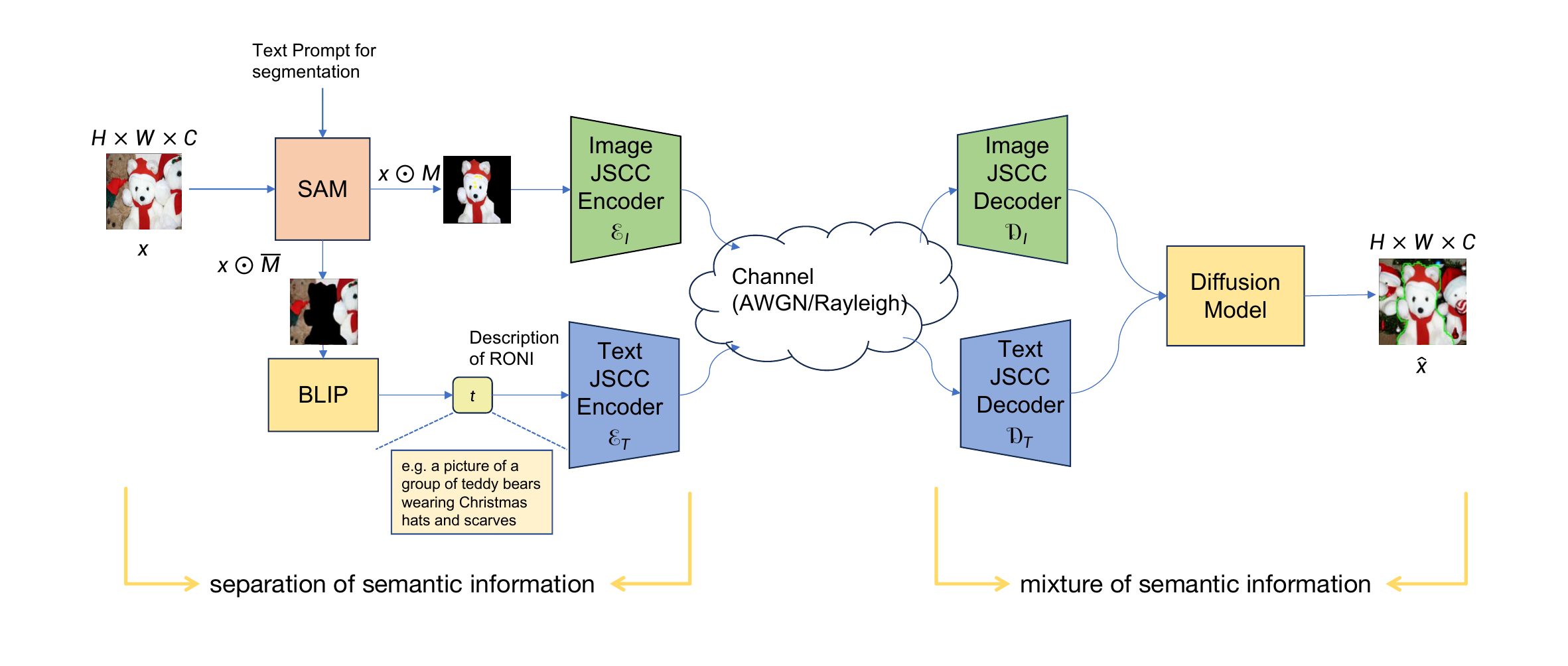}
	\caption{The proposed mixture of semantic transmission scheme. The foreground of received image is outlined in green.}
	\label{system}
	\vspace{-0.5cm}
\end{figure*}

The proposed MoS for wireless image semantic communications is shown in Fig. \ref{system}.


\subsection{Transmitter side}
We denote the input image as $\mathbf{x} \in \mathbb{R}^{H \times W \times 3}$, where $H, W$ represent the height and width of the input image. The transmitter consists of a segment model, a vision-language model and JSCC encoders for image and text respectively.

\subsubsection{Segmentation model}
The first step segments the image into ROI and RONI based on the user's prompt utilizating a segmentation model SAM, which consists of an Image Encoder, a Prompt Encoder and a Mask Decoder, all built upon Transformer-based vision models.
The SAM outputs a binary mask $\mathbf{M}$. Therefore, $\mathbf{x} \odot \mathbf{M}$ represents the ROI. Similarly, the RONI is obtained by $\mathbf{x} \odot \mathbf{\overline{M}}$. After separating the ROI and RONI, these parts are then processed parallelly.

\subsubsection{Image JSCC encoder}
The image JSCC encoder is responsible for the encoding the ROI of the image into symbols for transmission over the channel using an image JSCC encoder $\mathcal{E}_I$. This process is formulated as:
   \begin{equation}\label{first}
\mathbf{y_p} = \mathcal{E}_I(\mathbf{x} \odot \mathbf{M}),
   \end{equation} 
where $\mathbf{y_p} \in \mathbb{C}^{N \cdot \frac{C}{2}}$ is the image vector. Here, $N, C$ corresponds to the output dimension and channels of the encoder.

\subsubsection{Vision-language model and Text JSCC encoder}
For the RONI part of the input image, a vision-language model BLIP\cite{li_blip_2022} is used at the transmitter to convert it into text modality, denoted as $t$. This converted text serves as a prompt for reconstructing the original image at the receiver. Once converted to text, the corresponding Text JSCC encoder is applied to perform the encoding. The process is expressed as:
   \begin{equation}\label{second}
\mathbf{y_t} = \mathcal{E}_T(t) = \mathcal{E}_T(\texttt{BLIP}(\mathbf{x \odot \overline{M}})),
   \end{equation}
where $\mathbf{y_t} \in \mathbb{C} ^{l \cdot \frac{B}{2}} $ represents the text vector Here, $l$ is the maximum sequence length of the text, and $B$ is its dimension.

The image vector and the text vector are concatenated as $\mathbf{y}$ with power constraint $\mathbb{E}\left[\left|\left|\mathbf{y}\right|\right|_2^2\right]\leq P$, which is then transmitted through the typical additive white Gaussian noise(AWGN) or Rayleigh fading channels.

\subsection{Receiver side}
At the receiver, the signal is decoded into image of ROI and text of RONI. These are represented as $\mathcal{D}_p(\mathbf{\hat{y_p}})$ and $\mathcal{D}_T(\mathbf{\hat{y_t}})$, respectively, where $\mathbf{\hat{y_p}}$ and $\mathbf{\hat{y_t}}$ denote the noise-corrupted symbols. To recover the original image, a pretrained generative AI, such as Stable Diffusion for inpainting\cite{rombach_high-resolution_2022}, is utilized to integrate the mixture of semantic information and generate the final image. Stable Diffusion is a latent diffusion model (LDM) that operates in a compressed latent space for efficient text-to-image synthesis. It comprises: (1) a VAE that encodes images into latent representations $\mathbf{z} = \mathcal{E}(\mathbf{x})$ and decodes as $\mathbf{\hat{x}} = \mathcal{D}(\mathbf{z})$; (2) A U-Net denoiser $\epsilon_\theta(\mathbf{z_t}, t, \mathcal{E}(\mathcal{D}_I(\mathbf{\hat{y_p}})), \mathbf{M}', \mathbf{c}_{\text{text}})$ trained to predict noise in $\mathbf{z}_t$ conditioned on CLIP text embeddings $\mathbf{c}_{\text{text}} = \tau(\mathcal{D}_T(\mathbf{\hat{y_t}}))$. $\mathbf{M}'$ is downsampled mask; (3) A DDIM sampler that iteratively refines $\mathbf{z}_T \sim \mathcal{N}(0,\mathbf{I})$ to $\mathbf{z}_0$ via $\mathbf{z_{t-1}} = \sqrt{\alpha_{t-1}} \left( \frac{\mathbf{z}_t - \sqrt{1-\alpha_t} \epsilon_\theta(\cdot)}{\sqrt{\alpha_t}} \right) + \sqrt{1-\alpha_{t-1}} \epsilon_\theta(\cdot)$. The reconstructed image preserved the original ROI and features a RONI that is semantically similar to that of the original image.

\section{Design of Mixture of semantic Strategy}
In this section, we present a detailed description of the MoS strategy and the performance evaluate metrics.

\subsection{Strategy design}
In communications strategies utilizing generative AI, the conventional approach is to generate images that are either entirely identical at the pixel level or only semantically similar on a global scale. However. in practical, different regions of an image draw varying levels of attention from viewers. A generation method that enforces pixel-level consistency tends to lack flexibility, whereas one that relies solely on semantic similarity may compromise perceptual fidelity. The MoS is specifically designed to address this challenge by integrating flexibility with perceptual similarity, thereby optimizing the image generation strategy to meet practical requirements.

In fact, transmitting only the ROI portion of an image using JSCC enhances the quality of the reconstructed image at the receiver. In other words, when only the ROI is considered, fewer channel resources are needed to achieve comparable quality, which offers an opportunity for optimization. Under conditions where channel resources are severely limited (e.g., low SNR, low channel bandwidth ratio(CBR), or both) and the overall image quality would otherwise be extremely low, a strategy that transmits only the ROI can be adopted to preserve its quality. For the RONI, leveraging the capabilities of generative AI allows us to convert this portion into another modality—such as text—requiring fewer channel resources; the RONI is then reconstructed at the receiver based on this transformed modality. Moreover, because the semantic information from the ROI is integrated with that from the RONI during the generation process, this strategy reduces perceptual inconsistencies between the original and reconstructed images. Compared to generating an entirely new image, this approach offers increased controllability, ensuring a higher degree of similarity between the generated image and the original, while enabling users to flexibly allocate channel resources based on the specified ROI.

As shown in Algorithm \ref{alg:proposed}, the traning algorithm for the proposed MoS stratgey comprises five key steps: 1) distinction between ROI and RONI; 2) encoding the ROI; 3) performing a modality transformation; 4) encoding the RONI, and 5) integrating the semantic information to regenerate the image. 
For Step 1, the separation of the ROI and RONI lays the groundwork for subsequent modality transformation and independent encoding. Specifically, BLIP is employed to convert the RONI from image to text modality, thereby significantly reducing the number of symbols required for transmission and enabling more efficient channel resource allocation. After encoding, channel transmission, and decoding, the inpainting diffusion model at the receiver ensures that the regenerated image maintains semantic consistency with the text derived from the background.
\begin{algorithm}[t]
	\caption{Training algorithm of the MoS strategy}
	\label{alg:proposed}
	\KwIn{pretrained parameters of SAM, BLIP and Diffusion Model, training images $\mathcal{S}$, users' prompt $\mathcal{P}$}
	\KwOut{The trained new architecture $\mathcal{E}_T, \mathcal{E}_I, \mathcal{D}_T, \mathcal{D}_I$}
	load pretrained parameters of SAM, BLIP and Diffusion\;
	$\mathcal{I}_{\text{ROI}} \gets \emptyset$\;
	$\mathcal{T}_{\text{RONI}} \gets \emptyset$\;
	\For{$\mathbf{x,p}$ in $\mathcal{S,P}$} {
		$M \gets \texttt{SAM}(\mathbf{x, p})$ \;
		$\mathcal{I}_{\text{ROI}} \gets \mathcal{I}_{\text{ROI}} \cup \{\mathbf{x \odot M} \} $ \;
		$\mathcal{T}_{\text{RONI}} \gets \mathcal{T}_{\text{RONI}} \cup \{\texttt{BLIP}(\mathbf{x \odot \overline{M}})\}$
	}
	\Repeat{JSCC converged}{
		Train image JSCC on $\mathcal{I}_{\text{ROI}}$\;
		Train text JSCC on $\mathcal{T}_{\text{RONI}}$\;
	}
\end{algorithm}

\begin{minipage}[htbp]{0.48\textwidth}
\centering
\captionof{table}{The selection of $\alpha$ under different channel conditions (Rayleigh)}
\resizebox{0.9\textwidth}{!}{
\begin{tabular}{cccccc}
    \toprule 
    SNR(dB)/CBR & 0.016 & 0.024 & 0.032 & 0.04 & 0.048 \\ 
    \midrule                      
    0       & 24.41\% & 16.28\% & 12.21\% & 9.77\% & 8.14\% \\
	  5       & 8.39\%  & 5.59\% & 4.20\% & 3.36\% & 2.80\% \\
	  10      & 6.10\%  & 4.07\% & 3.05\% & 2.44\% & 2.03\% \\
	  15      & 4.58\%  & 3.05\% & 2.29\% & 1.83\% & 1.53\% \\
	  20      & 3.81\%  & 2.54\% & 1.91\% & 1.53\% & 1.27\% \\
    \bottomrule                     
  \end{tabular}
}
\label{tab:alpha_Rayleigh}
\end{minipage}

\subsection{Performance evaluate metrics}
For evaluation metrics, we divide each image into the ROI and RONI regions for separate assessments. For the ROI, where pixel-level consistency is critical, we employ the PSNR-ROI as the evaluation metric.

For the RONI, the goal is to assess semantic-level similarity. To this end, we employ the CLIP Score\cite{hessel2022clipscorereferencefreeevaluationmetric} as the evaluation metric, which is widely used to assess image-text consistency. Since only RONI part of image is considered, we can also call it CLIP-RONI. The metric is defined as
\begin{equation}
\text{CLIP-RONI}(\mathbf{c},\mathbf{v}) = \max\left(\cos(\mathbf{c},\mathbf{v}),0\right),
\end{equation}
where $\mathbf{c},\mathbf{v}$ denote the the visual CLIP embedding of the image in RONI and the textual CLIP embedding of the text, respectively. 

\section{Simulation Results}
\label{sim}
\subsection{Experiment Setup}
In the experiments, we use the RefCOCO dataset. The RefCOCO dataset is a subset of COCO designed for referring expression generation (REG) tasks, whose objective is to understand natural language expressions that refer to specific objects in images. We employ the descriptions corresponding to specific image regions as prompts for segmentation. The training set comprises 15,786 image, object and description pairs, while the testing set includes 1,810 pairs. Furthermore, we also use Pascal VOC datasets for expanding the diversity. During training, each image was resized to $256 \times 256 \times 3$.

We incorporated the baseline method SwinJSCC, DiffCom\cite{DiffCom} and ROI-JSCC\cite{roijscc} as comparative benchmarks in our analysis to provide a comprehensive evaluation framework. DiffCom is a generative communication paradigm that uses probabilistic diffusion models for decoding. ROI-JSCC is a novel deepJSCC framework that prioritizes high-quality transmission of ROI to selectively improve the reconstruction quality of varying ROI while preserving overall image quality. The DiffCom strategy employs ADJSCC for its JSCC component. The diffusion model was trained on the ImageNet dataset. For ROIJSCC, the encoder and decoder are configured with the default settings from the original paper, which involves 4 stages, each composed of 2 ROI blocks.

We implement a image JSCC encoder based on the Swin Transformer with a block configuration of $[2, 2, 2, 2]$. The receiver configuration is identical to that of the transmitter. The channel SNR varies between 0 to 20 dB and the CBR varies from 0.01 to 0.05. For text transmission, we adopte a JSCC encoder based on the transformer, which comprises 4 transformer blocks. The image JSCC is trained and evaluated by Adam optimizer with a learning rate of $10^{-4}$ over 200 epochs, while the text JSCC is trained for 100 epochs.

\subsection{Result Analysis}
\label{res analysis}

\begin{figure}[t]
	\centering
	\includegraphics[width=0.35\textwidth]{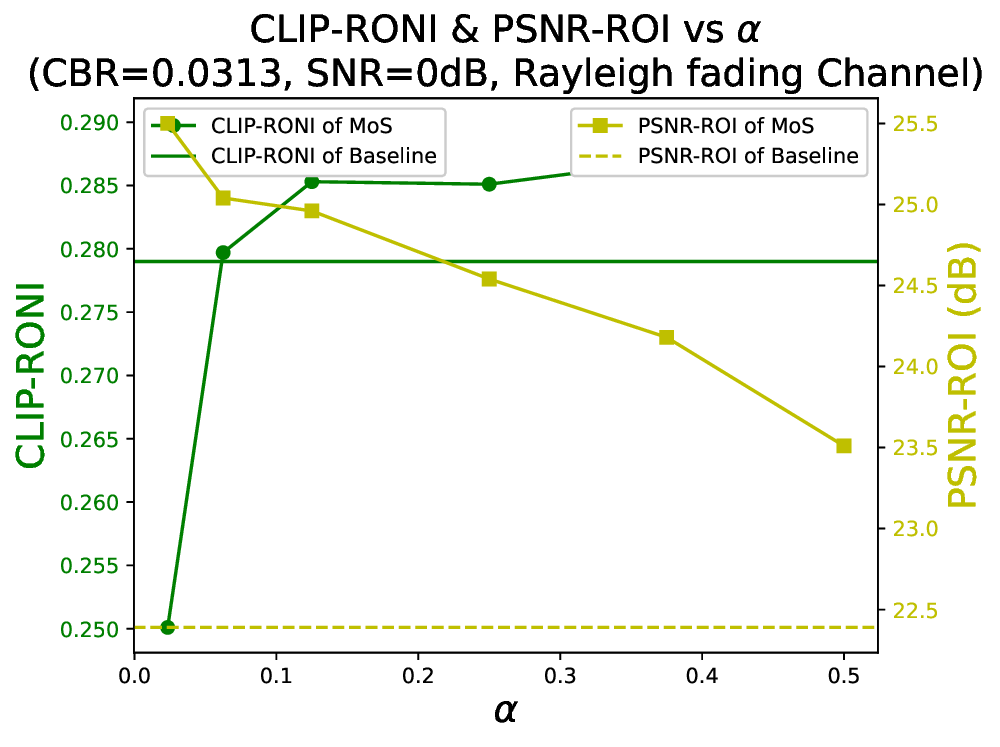}
	\caption{Impact of $\alpha$ on system performance.}
	\label{balance}
	\vspace{-0.2cm}
\end{figure}

Both PSNR-ROI and CLIP-RONI metrics are influenced by the allocation of symbols between image and text. To quantify the influence, we introduce a key hyperparameter in the experiment: the proportion of symbols allocated to represent text (or RONI) among all symbols, denoted as $\alpha$.

\begin{figure}[t]
	\centering
	\includegraphics[width=0.48\textwidth]{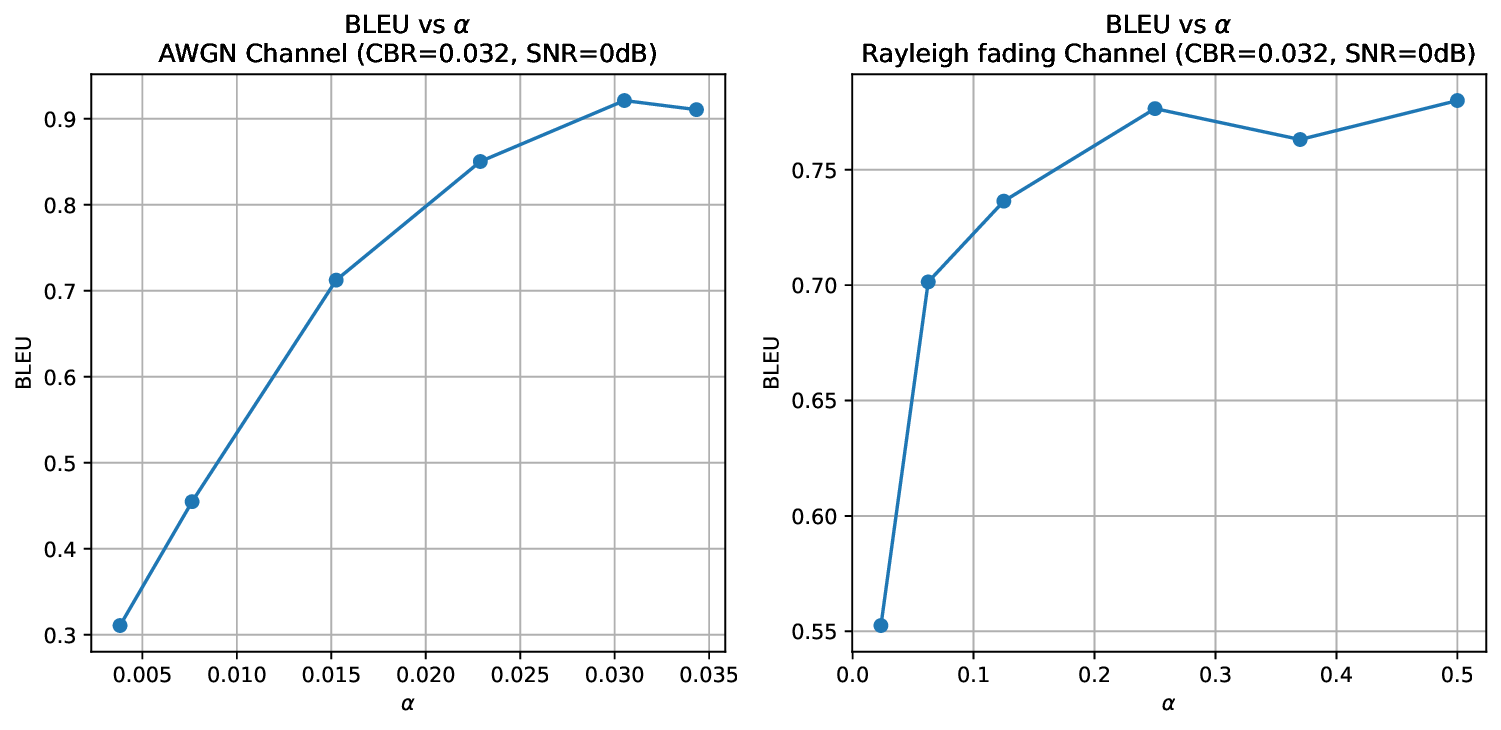}
	\caption{BLEU versus $\alpha$ at SNR= 0 dB and CBR=0.032}
	\label{fig:bleua}
	\vspace{-0.5cm}
\end{figure}

Fig. \ref{balance} illustrates the impact of $\alpha$ on system performance under Rayleigh fading channels at an SNR of $0$ dB and a CBR of $0.03$.
We find that the CLIP-RONI Score exhibits a threshold effect, which means that its growth slows down even stop at 0.285 when $\alpha$ reaches the thereshold and it decreases fastly when $\alpha$ is lower than the thereshold. This is because the CLIP-RONI Score performance is based on the gerneration content. Only text is utilized to guide the gerneration, whose error rate versus $\alpha$ is shown in Fig. \ref{fig:bleua}. Due to the robustness of the gerneration model, a slight error in text does not importantly impact the gerneration content but a great error will significantly misguide the gernerative model. Therefore, increasing $\alpha$ to allocate more resources the text transmission bring little performance gain.
However, increasing $\alpha$ means allocates fewer resources for transmission ROI, causing further decreasing in the PSNR-ROI. Therefore, we set the $\alpha$ to the thereshold point of CLIP-RONI Score.
The selection of $\alpha$ over the Rayleigh fading channel is shown in Table \ref{tab:alpha_Rayleigh}. 

\begin{figure}[t]
	\centering
	\includegraphics[width=0.485\textwidth]{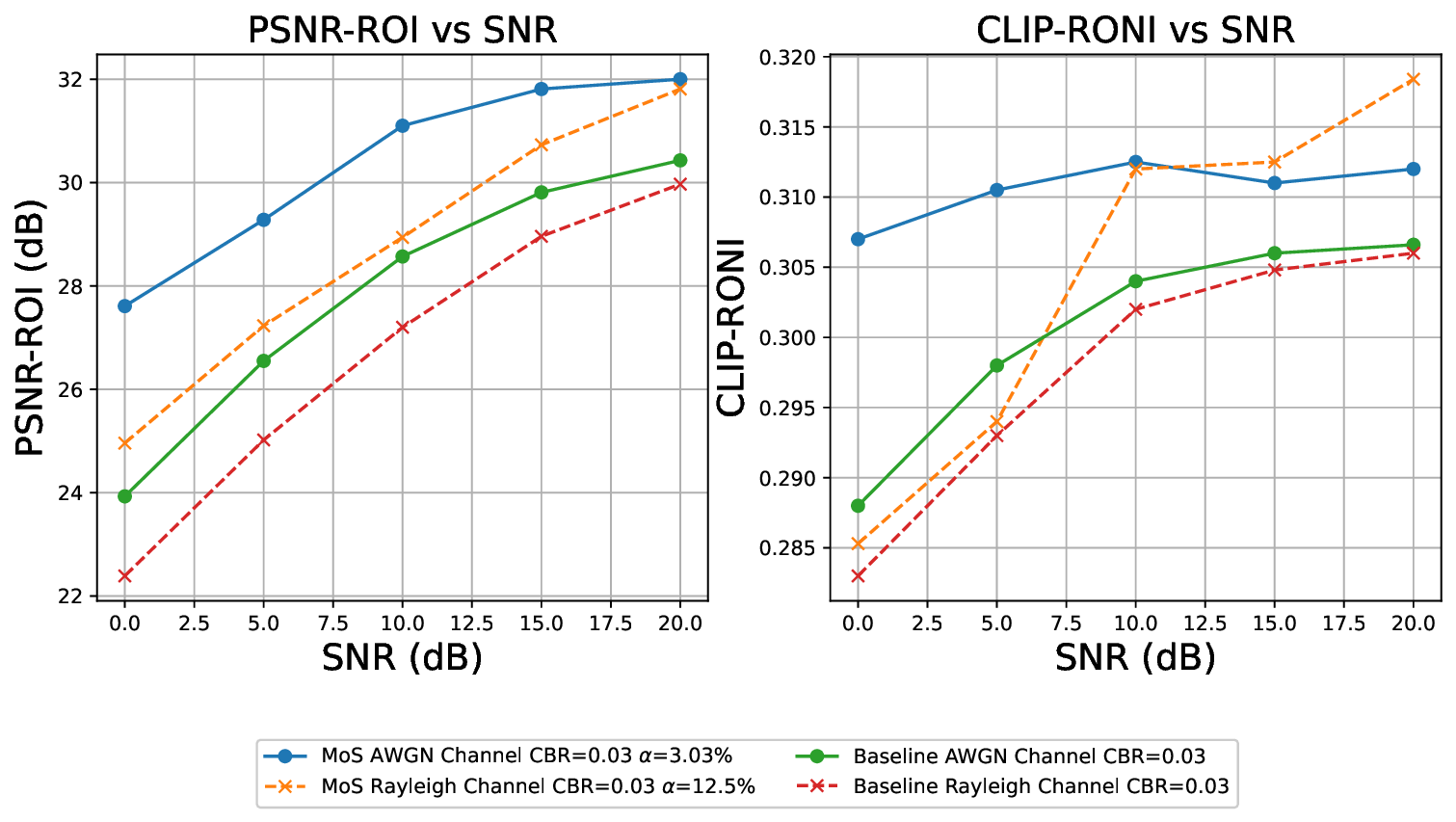}
	\caption{
	The left figure presents PSNR-ROI versus SNR and the right presents CLIP-RONI versus SNR. 
	}
	\label{snr}
\end{figure}

\begin{figure}[t]
	\centering
	\includegraphics[width=0.45\textwidth]{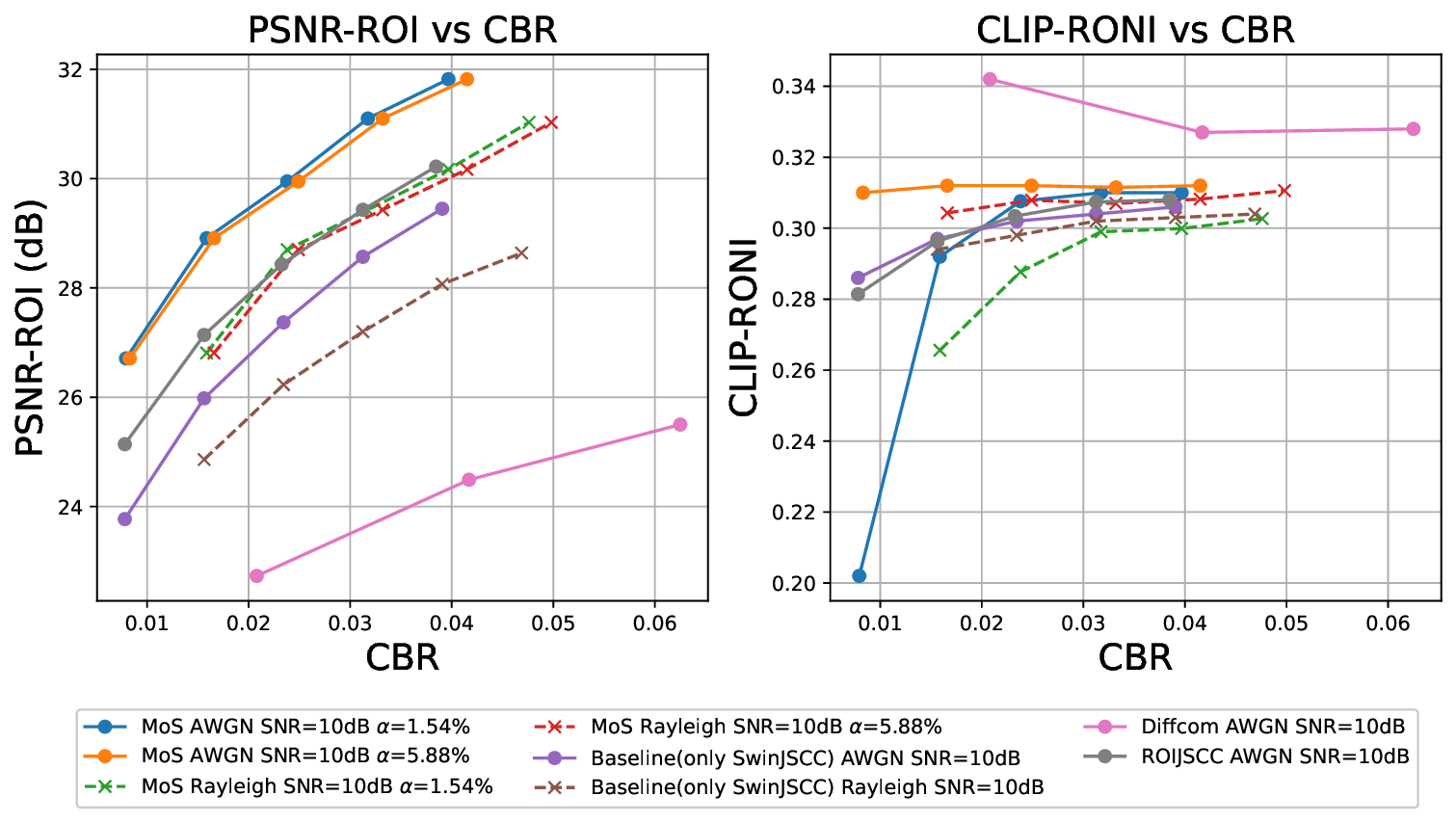}
	\caption{
	The left figure presents PSNR-ROI versus CBR.
	The right figure presents CLIP-RONI Score versus CBR.}
	\label{cbr}
	\vspace{-0.5cm}
\end{figure}

Fig. \ref{snr} and Fig. \ref{cbr} presents the PSNR-ROI and CLIP-RONI performance of the proposed MoS compared to the baseline under both AWGN and Rayleigh fading channels. Fig.\ref{snr} (left) illustrates the performance at a fixed CBR of 0.03, the proposed MoS achieves a PSNR improvement of approximately $2$ to $4$ dB over the baseline across multiple SNR levels, e.g. a PSNR-ROI gain of 3.68dB PSNR-ROI over baseline under 0dB, AWGN channel. In Fig. \ref{snr} (right), across various SNR levels, the proposed MoS achieves a CLIP-RONI score improvement ranging from 0.001 to 0.02 compared to the baseline,e.g. a CLIP-RONI gain of 0.019 over baseline under 0dB, AWGN channel. The results demonstrate that the proposed MoS consistently outperforms the baseline across various SNRs.
\begin{figure}[t]
	\centering
	\includegraphics[width=0.45\textwidth]{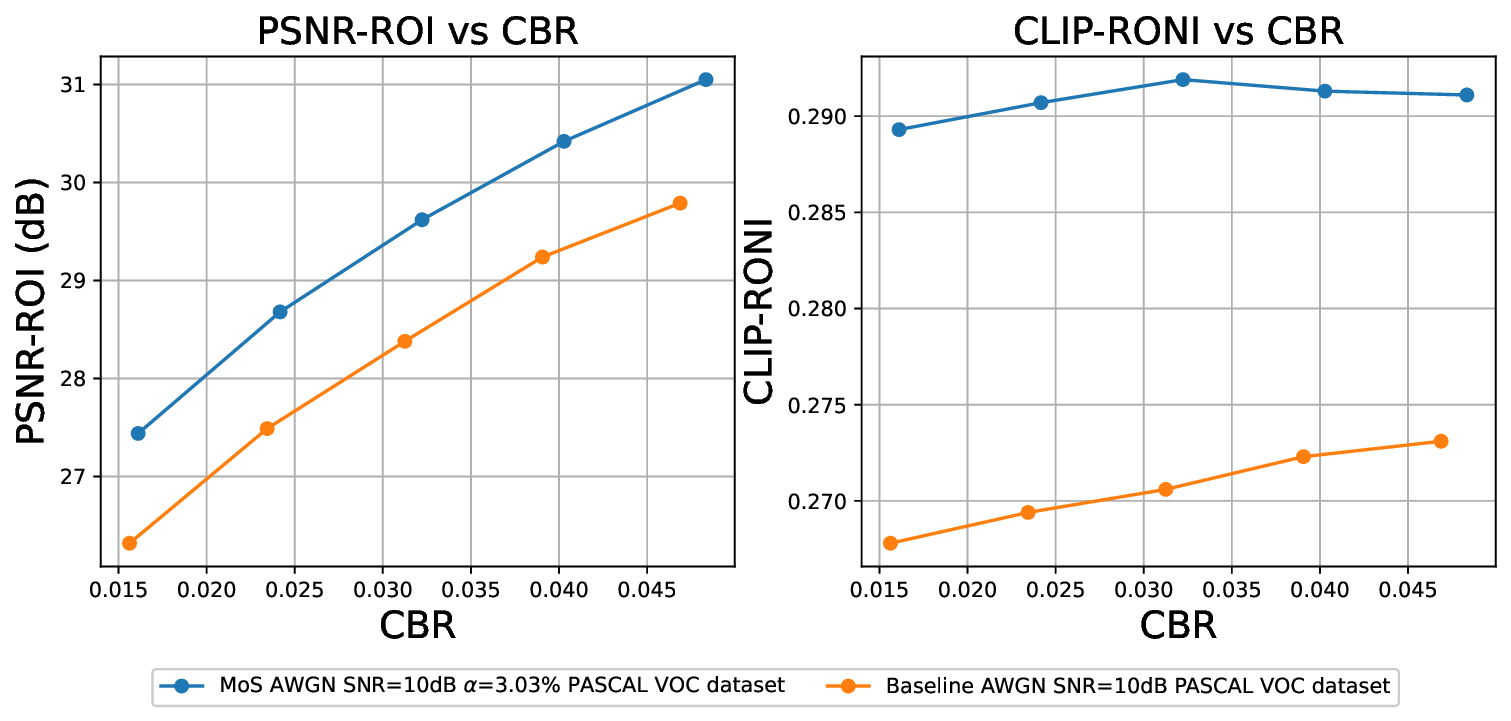}
	\caption{Results on Pascal VOC dataset}
	\label{fig:VOC}
\end{figure}

In Fig. \ref{cbr} (left), with the SNR of 10 dB, the proposed MoS exhibits a PSNR-ROI gain of $2-3$ dB over the baseline across different CBR levels. Furthermore, in both Rayleigh and AWGN channels, increasing $\alpha$ from $1.54\%$ to $5.88\%$ leads to a slight decline in PSNR-ROI. However, as shown in Fig. \ref{cbr} (right), the increase of $\alpha$ makes the CLIP-RONI score surpasses the baseline, aligning with the trends observed in Fig. \ref{balance}, proving the importance of the selection of $\alpha$. Our MoS strategy also  outperforms in PSNR-ROI and CLIP-RONI score than ROI-JSCC and better PSNR-ROI performance than DiffCom. However, our MoS strategy performs worse than DiffCom on CLIP-RONI because the DiffCom allocates more resources for transmission the RONI.

Apart from RefCOCO dataset, we also conduct our experiments on Pascal VOC dataset. As shown in Fig. \ref{fig:VOC}, it demonstrates the same trend of PSNR-ROI and CLIP-RONI as experiments performed on RefCOCO dataset, which prove that our MoS preserves performance on diverse datasets.

\begin{table}[htbp]
	\vspace{0.2cm}
  \centering
  \caption{Computational complexity, parameter size and latency implications comparison}
  \begin{tabular}{llll}
    \toprule                       
      & \# of Parameters & FLOPs & latency \\
    \midrule                    
	  MoS & 2.05B & 19.08T & 2.588s \\   
    Baseline & 0.02B & 27.81G & 42.819ms\\
    DiffCom & 0.55B & 1449.24T & 33.67s\\
    \bottomrule               
  \end{tabular}
  \label{tab:computational complexity and latency}  
\end{table}

\begin{figure}[t]
	\centering
	\includegraphics[width=0.48\textwidth]{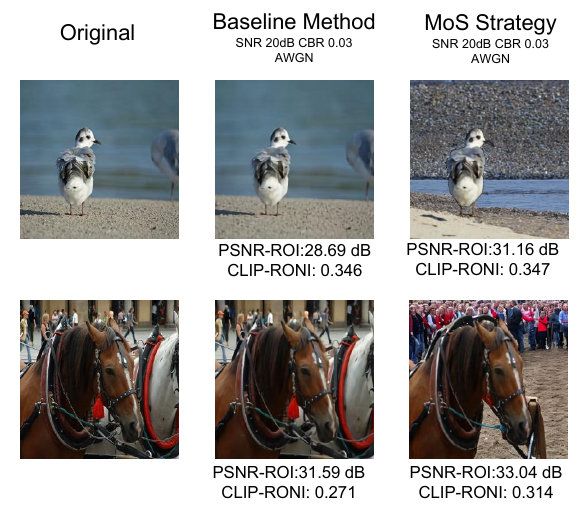}
	\caption{Visual results: the original images (left), the baseline method (middle), and the proposed MoS method (right).}
	\label{fig:comparison}
	\vspace{-0.5cm}
\end{figure}

Table \ref{tab:computational complexity and latency} shows the computational complexity, latency implications and parameter size of our MoS strategy, baseline scenario and DiffCom.  
From the table, it can be observed that due to the complex architecture for achieving remarkable gains with generative models, our method inevitably introduces much computational overhead and parameters  compared with the non-generative baseline method. 
However, compared to the generative model based DiffCom method, though our MoS strategy still requires more parameters, its computational overhead and interference latency is significantly lower because we consider that the inference latency is important for real-time communication and thus have adopted an advanced sampling method to accelerate the generation process.
Fig. \ref{fig:comparison} presents visual results. The SNR is set to 20 dB, and CBR 0.03 under AWGN channel. It is evident that our proposed MoS method typically yields both higher PSNR-ROI and higher or comparable CLIP-RONI scores than the baseline method. 

\section{Conclusion}
\label{con}
In this paper, we proposed a novel communication strategy—MoS, which leverages modality decomposition and generative AI for wireless image transmission. The proposed MoS approach divides an image into distinct components, applies modality-specific encoding, and regenerates the image at the receiver. This method reduces the discrepancies between the transmitted and received images compared to other generative AI-based techniques while offering enhanced flexibility in channel resource allocation. Our findings reveal that balancing the proportions of text and image symbols is crucial, particularly under challenging channel conditions. Extensive experiments demonstrate significant improvements in both PSNR-ROI and CLIP-RONI metrics.

\bibliography{reference}
	
\end{document}